\documentclass[aps,twocolumn,amsmath,floatfix,byrevtex,prb,nobibnotes]{revtex4}
\usepackage{graphicx}
\usepackage{amssymb}
\usepackage{bm}

\begin{document}

\begin{abstract}

Electron counting of a single porphyrin molecule between two electrodes shows a crossover from sub- to super-Poissonian statistics as the bias voltage is scanned. This is attributed to the simultaneous activation of states with electron transfer rates spanning several orders of magnitude. Time-series analysis of consecutive single electron transfer events reveals fast and slow transport channels, which are not resolved by the average current alone.

\end{abstract}

\title{Single-electron counting spectroscopy: simulation study of porphyrin in a molecular junction}

\author{Sven Welack}
\thanks{also at: Department of Chemistry, Hong Kong University of Science and Technology, Kowloon, Hong Kong}
\email{swelack@uci.edu}
\author{Jeremy B. Maddox}
\thanks{present address: Department of Chemistry and Biochemistry, Texas Tech University, P.O. Box 41061, Lubbock, Texas 794091-1061}
\author{Massimiliano Esposito}
\thanks{present address: Department of Chemistry and Biochemistry University of California San Diego, La Jolla, CA 92093, USA and Center for Nonlinear Phenomena and Complex Systems, Universite Libre de Bruxelles, Code Postal 231, Campus Plaine, B-1050 Brussels, Belgium}
\author{Upendra Harbola}
\author{Shaul Mukamel}\email{smukamel@uci.edu}
\affiliation{Department of Chemistry, University of California, Irvine, California 92697, USA.}

\maketitle

Real-time measurements of single-electron transfers through nanosystems have been reported recently\cite{Lu03,Fuji04,Byla04,Gustav06,Fuji06,Gruneis07}. Although electron counting has not yet been observed in single molecule junctions, new techniques based on carbon nanotubes have been proposed for its possible realization\cite{Gruneis07}. Theoretical efforts are required to fully connect the information of the electron transfer statistics to the molecular properties. This will offer new opportunities to study many-body effect to characterize the bonding of molecules to external electrodes with implications to molecular electronics\cite{nitz03a,ghos04}.

Full electron counting statistics \cite{levi96,Bagr03,Levi04,Blant00,Gurv97,Wabnig05,Rammer04,Shela03,Flindt05,Kiess06,Peder05,groth07,max06a,sven07_2,fleurov07} provides detailed information about the probability $P(k,t)$ of transferring a net-number $k=k_{in}-k_{out}$ of electrons between electrode and molecule during a time interval $t$. The cumulants of $P(k,t)$ are directly related to important properties of the junction. For example, the first order cumulant $C_1(t)=\bar k / t$ gives the average current $I(t)=e C_1(t)$. The shot noise is related to the second cumulant $S(t)=2 e^2 C_2(t)=2 e^2 (\overline{k^2} - {\bar k}^2)/t$ and is commonly represented by the Fano factor $F(t)=C_2(t)/C_1(t)$. The third cumulant $C_3(t)=\overline{(k-\overline{k})^3}/t$ measures the skewness of the distribution.

\begin{figure}
\includegraphics[width=7.0cm,clip]{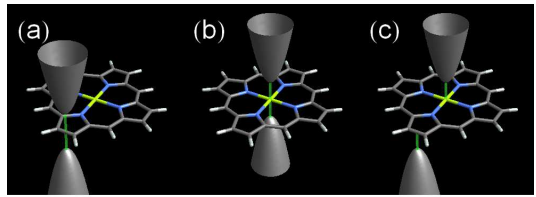}
\caption{The geometrical configurations (a),(b) and (c) of the source (top) and drain (bottom) electrode with respect to the MgP molecule.
\label{fig:1}}
\end{figure}

The goal of the present work is to relate the higher statistical cumulants to the intrinsic properties of the metal-molecule interface which cannot be accessed by average current measurements. We simulate a magnesium porphine (MgP) molecule coupled to two metallic electrodes as shown in Fig.\,\ref{fig:1} and find that the Fano factor shows a crossover from sub- ($F<1$) to super-Poissonian ($F>1$) shot noise when a large number of transport channels are opened by the bias. 

Mechanisms like asymmetric coupling strengths of the two electrodes\cite{thiel03,thiel05,wang07,belzig05}, Coulomb charging energy in multi-level systems\cite{belzig05,thiel03} and potential fluctuations in resonant quantum wells\cite{Buett05} were predicted to induce super-Poissonian transfer statistics. In case the two electrodes are coupled to the molecule with equal strength, we identify an additional mechanism in the MgP-metal bonding and attribute it to the presence of couplings between the electronic many-body states of the molecule spanning different orders of magnitude. This is verified using a simple model.

The presence of these overlapping effects makes it difficult to analyze specific electronic states and couplings using the statistical cumulants. We further examine the probabilities of consecutive electron transfers which can be obtained from a time-series of single electron transfer events\cite{sven07_2} and calculate a decay time distribution of the MgP molecule junction. We demonstrate that slow and fast transport channels can be unambiguously identified as peaks in the decay time distribution.

The junction Hamiltonian is decomposed as $H=H_0+H_{S}+H_{D}+H_T$, where $H_0$ represents the isolated MgP molecule, $H_{S(D)}$ the source (drain) electrode, and $H_T$ is the electrode-molecule interaction. We assume that the density of states of the electrode is constant on the relevant energy scale and that the current is small.

Describing the electrodes as free electron reservoirs and neglecting coherence elements, the Pauli rate equation provides a convenient framework to include many-body effects\cite{Datta06,Schoell03} and electronic structure calculations\cite{Elste05}. The Pauli equation for the population $p_i^{(N)}$ of state $\vert i \rangle$, eigenstate of $H_0$, with energy $E_i$ and $N$ electrons is then given by
\begin{equation}\label{equ:rate1}
\dot{p}_{i}^{(N)} = \sum_{j (\neq i)} \big( W_{j \to i} p_{j}^{(N \pm 1)} - W_{i \to j} p_{i}^{(N)} \big).
\end{equation}
The molecule can be excited by electron transfers from the electrodes. Only transitions between states with a unit charge difference are allowed. The transition rates $W_{i\to j}$ involve contributions from source ($\alpha=S$) and drain ($\alpha=D$) electrode and may be decomposed as
$
W_{i\to j}^{(\pm)}=W_{i\to j}^{(S,\pm)}+W_{i\to j}^{(D,\pm)}
$
where 
$W_{i\to j}^{(\alpha,+)}=\Gamma_{ji}^{(\alpha)}f_\alpha(E_{ji})$ 
is the rate of electron transfer into the molecule and 
$W_{i\to j}^{(\alpha,-)}=\Gamma_{ji}^{(\alpha)}(1-f_\alpha(E_{ij}))$
accounts for the reverse process. $E_{ij}=E_i-E_j$
is an energy difference between charge states 
and $f_\alpha(E)=(1+\exp [\beta(E-\mu_\alpha)])^{-1}$
is the Fermi function for a free electron gas with thermal energy $1/\beta$
and chemical potential $\mu_\alpha$. Each pair $W_{i\to j}^{\pm}$ of transition rates defines a transport channel from the source to the drain which can be opened and closed by the chemical potentials. The transition rate between a pair of states depends on the electrode-molecule 
interaction through the couplings 
$\Gamma_{ij}^{(\alpha)}=\sum_{s}\vert T_s^{(\alpha)}\vert^2\vert\langle j\vert c_s\vert i\rangle\vert^2$
where the index $s$ runs over the molecular orbitals and 
$T_s^{(\alpha)}$ is a coupling strength parameter weighted by the 
overlap $\langle j\vert c_s\vert i\rangle$ between many-body states. 

\begin{figure}
\includegraphics[width=7.0cm,clip]{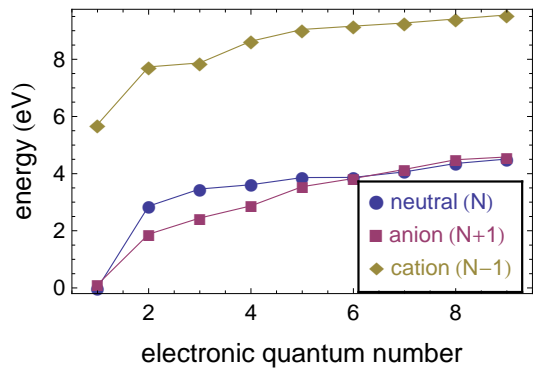}
\caption{The eigenenergies of the electronic states for the neutral, anionic and cationic charge states of the MgP molecule. 
\label{fig:1c}}
\end{figure}

Ground and excited states of the neutral and the charged MgP are calculated at the Hartree Fock (HF) and configuration interaction singles (CIS) level using a 6-31G basis set\cite{g03}. We fix the molecular geometry of the minimum energy configuration of the neutral MgP. Fig.\,\ref{fig:1c}  shows the nine lowest electronic energies of the neutral $N$, anionic $N+1$ and cationic $N-1$ MgP charge states; the neutral molecule's ground state energy is taken to be zero. The connecting lines are a guide for the eye. These states are connected by a network of couplings $\Gamma^{(\alpha)}_{ij}$.  For states $\vert i \rangle$ and $\vert j\rangle$ the overlap factor $\langle j |c_s|i \rangle$ is calculated using the CIS expansion coefficients, the molecular orbital coefficients, and the atomic overlap matrix; the procedure is described elsewhere\cite{maddox07}. The overlap factors for MgP show that multiple orbitals contribute to the electron transfer rate indicating that a many-electron picture is necessary to describe the transport.  These are shown in the supplementary information section.

\begin{figure}
\includegraphics[width=7.0cm,clip]{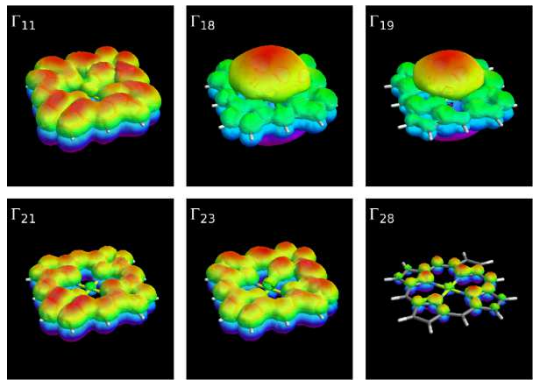}
\caption{The spatial profiles of the couplings $\Gamma_{ij}^{(\alpha)}$ between the electronic states $(i,N)$ and $(j,N+1)$.
\label{fig:1b}}
\end{figure}

The $T_s^{(\alpha)}$ factors are assumed to take the form of a kinetic energy integral between molecular orbital $\vert s\rangle$ and a spherically symmetric probe orbital $\vert\alpha\rangle$ centered about a point $r_\alpha$ representing the lead-molecule contact. The probe orbitals used in our calculations are centered at the tips of the gray paraboloids shown in Fig.\,\ref{fig:1}.  A useful representation of the coupling is obtained by assuming the probes have perfect spatial resolution, {\em i.e.},  $\langle r\vert\alpha\rangle=\delta(r-r_\alpha)$.  In this case $T_s^{(\alpha)}\propto\nabla^2\phi_s(r_\alpha)$ where $\phi_s(r_\alpha)$ is the orbital wavefunction evaluated at the probe's center such that the coupling $\Gamma^{\alpha}_{ij}$ can be readily visualized in 3D space. In Fig.\,\ref{fig:1b}, we show the $\Gamma_{11}$, $\Gamma_{18}$, $\Gamma_{19}$, $\Gamma_{21}$, $\Gamma_{23}$ and $\Gamma_{28}$ couplings between the neutral and anionic charge state plotted with respect to a delta-function probe's central position. The couplings are highly sensitive to the electrode-molecule orientation.

The steady state is calculated by setting $\dot{p}_{i}^{(N)} =0$ in Eq.\,(\ref{equ:rate1}). From the resulting homogeneous system of equations an inhomogeneous form can be derived using the additional constraint $\sum_i p_i=1$. For an irreducible rate matrix, the inhomogeneous system can be solved algebraically for $p$. 

We use a generating function technique to calculate the electron transfer statistics through the junction\cite{levi96,Bagr03,Levi04,Blant00,Gurv97,Wabnig05,Rammer04,Shela03,Flindt05,Kiess06,Peder05,groth07,max06a,sven07_2}. The asymptotic cumulants, describing an infinite time counting measurement, can be calculated from the eigenvalues of the generating function propagator\cite{max06a,sven07_2}. For details see supplementary material. The statistics is generated by counting the net-transfer (in minus out) at the source. Since $\beta V>1$ in our calculations, electron transfers in opposite direction of the bias are insignificant and the net-process is practically reduced to transfers with the bias.

We employ the electrode configurations (a),(b) and (c) shown in Fig.\,\ref{fig:1}. The probe orbitals are centered at the tips of the gray paraboloids. The upper electrode acts as the source, the bottom is the drain. The chemical potential of the source is fixed at $\mu_S=E_F=2.0eV$ and the drain chemical potential $\mu_D=E_F-V$ is decreased. The voltage is varied in the energy range of the employed states between $0$ and $4.75 V$. In this range we can safely limit our discussion to the neutral and anion states since the cationic states are inaccessible. The temperature is set to $T=50 K$. The current is given in nano Ampere (nA) and time in nano seconds (ns).

\begin{figure}
\includegraphics[width=7.0cm,clip]{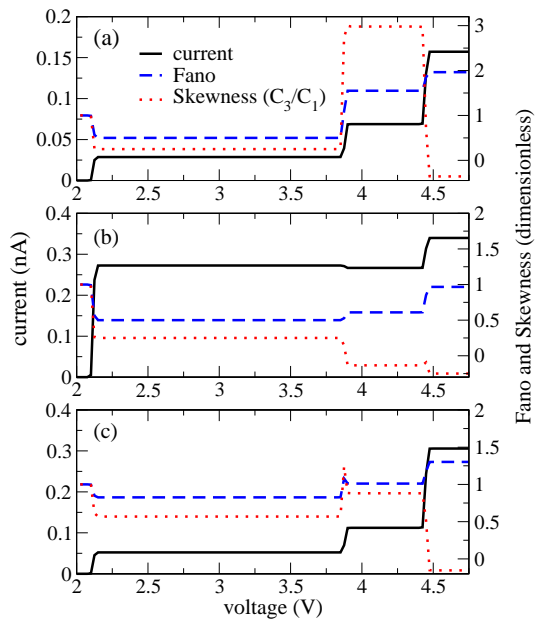}
\caption{The net-current through the MgP junction as a function of voltage for electrode geometries (a), (b) and (c). 
The corresponding Fano factor and the skewness of the electron transfer statistics at the source electrode are quantified by the right scale. 
\label{fig:2}}
\end{figure}

In configuration (a) both electrodes are symmetrically located at carbon ring. The corresponding current-voltage curve (Fig.\,\ref{fig:2}a) shows three steps when charge excitations in the molecule are energetically allowed. Since the equilibrium Fermi energy is set to $E_F=2.0 eV$, the first transition, between the neutral ground state $N, i=1$ and the lowest anionic state $N+1, i=1$, is shifted by $2.0 eV$ as well. Below a bias of $3.7 eV$ only these two states contribute to the current. Higher bias causes a cascade of excitations and additional states become occupied. The population dynamics is shown in the supplementary material.


Interesting features can be observed in the higher cumulants of the electron transfer statistics through the contact. Fig.\,\ref{fig:2} shows the Fano factor and skewness $C_3/C_2$ on the right scale. For small bias, the current is small and the single electron transfers are uncorrelated leading to a Poissonian process with Fano factor $F=1$. It drops to a sub-Poissonian regime $F<1$ when only the $\Gamma_{11}$ coupling is active. However, it becomes super-Poissonian $F>1$ at higher bias. The skewness is even more sensitive and reaches negative values in the super-Poissonian regime at $V>3.8 eV$. 

It has been shown\cite{thiel03,thiel05,wang07,belzig05} that a strongly asymmetric coupling of the system to the source and drain electrode can cause super-Poissonian electron transfer. This can be ruled out since the junction is symmetrically coupled to drain and source. We find that the sub-to super-Poissonian crossover can be attributed to two mechanisms. One is well known due to Coulomb charging energy in multi-level systems\cite{belzig05,thiel03}. 

The second effect can be attributed to large or small ratios of active excitation rates $W_{i \to j}^{(\alpha )}/W_{i\to l}^{(\alpha )}$ from a given (N,i) state to two (N+1,j),(N+1,l). Because $\beta$ is much smaller than the energy gaps between the electronic states, the Fermi function approximates a heavy-side step function and the ratio of the rates is directly related to the the corresponding couplings $\Gamma_{i \to j}/\Gamma_{i \to l}$.

This generic effect is rationalized in the supplementary material using a simple model. For ratios of the couplings of the same electrode larger than $5$ or smaller than $1/5$ we observe super-Poissonian electron transfer statistics. Since it does not depend on an asymmetric coupling between source and drain it can be considered as an intrinsic characteristic of the molecule-electrode interface. For example in Fig.\,\ref{fig:1b} we see that $\Gamma_{18}^{(\alpha)}$ coupling is weak compared to $\Gamma_{11}^{(\alpha)}$ at the edge of MgP and we find a ratio of $\Gamma_{11}^{(\alpha)}/\Gamma_{18}^{(\alpha)} \approx 10$ at the position of the tip.

We next move both electrodes to the center of MgP as shown in Fig.\,\ref{fig:1} (b). The current, Fano factor and skewness are shown in Fig.\,\ref{fig:2}b. Again for a bias below $3.7V$ only the $W_{1\to 1}^{\pm}$ channel is active and the electron transfer is sub-Poissonian as indicated by the Fano factor. Once the additional channels are opened by the bias the Fano factor increases but does not exceed unity since the coupling ratio of the most significant channels is within the limits for sub-Poissonian statistics as predicted by our model calculations. In the center of the MgP, $\Gamma_{18}^{(\alpha)}$ is strong as shown in Fig Fig.\,\ref{fig:1b} and we have a ratio of $\Gamma_{11}^{(\alpha)}/\Gamma_{18}^{(\alpha)}\approx 1/4$ at the tip. Here the increase of the Fano factor is due to the Coulomb charging and this electrode configuration provides a good quantitative estimate for it.

Let us move the drain back to its position in configuration (a) while the source is kept in the center as shown in Fig.\,\ref{fig:1} (c). Besides the Coulomb charging and extreme ratios of the couplings, we now also have asymmetric coupling to drain and source and three possible sources for super-Poissonian transfer are present. However the spatial asymmetry can be estimated from the average current signal of the two symmetric configurations. The Fano factor in Fig.\,\ref{fig:2}c shows the crossover but is with a maximal value of $F=1.26$ smaller than in configuration (a). Qualitatively, current signal, Fano factor and skewness resemble an average over the corresponding signals of (a) and (b). The asymmetry effect is therefore weaker in configuration (c) than the other two.

\begin{figure}
\includegraphics[width=7.0cm,clip]{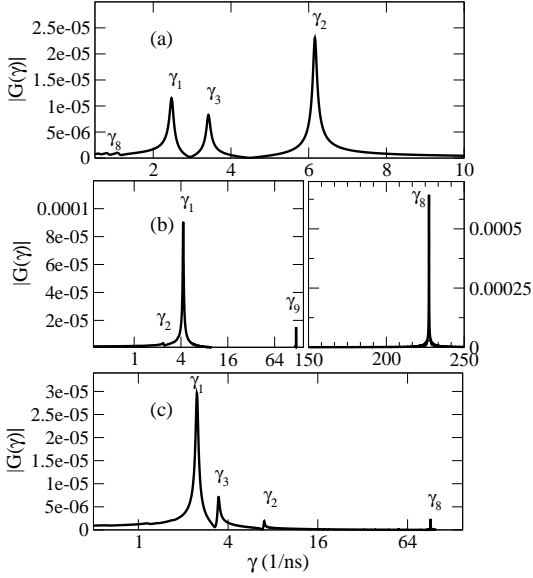}
\caption{The decay time distribution $G_{S \rightarrow D}(\gamma)$ for the electrode geometries (a)-(c). A finite width of the peaks was introduced by modulating the signal with a frequency of $\omega=7.5 10^4 1/ns$. A logarithmic $\gamma$-axis is used for the left side of the middle panel and the bottom panel. 
\label{fig:5b}}
\end{figure}

We next examine the elementary probabilities of consecutive electron transfer events\cite{gardiner,Muka03,sven07_2}. An electron is detected when it enters the junction through the source at time $t_0=0$ and leaves through the drain at time $t$. No other transfer events take place in between. We denote this path by $S \rightarrow D$. The two corresponding electron transfer operators are $W^{(S,+)}$ and $W^{(D,-)}$ respectively. Their matrix elements are given by the transfer elements of the Pauli equation (\ref{equ:rate1}). The elementary probability of this two-time measurement is then given by\cite{sven07_2}
\begin{equation}\label{lr1}
P_{S \rightarrow D}(t, t_0) = \langle W^{(D,-)}  U(t,t_{0}) W^{(S,+)} \rangle,
\end{equation}
where $U(t,t_0)$ is the propagator of the system in absence of transfer events at the electrodes within the time interval
$
U(t,t_0)=e^{\sum_{\alpha=S,D} -W^{(\alpha,diag)} (t-t_0) }
$
The operator $W^{(\alpha,diag)}$ contains diagonal elements of the Pauli rate matrix calculated by $W^{(\alpha,diag)}=\sum_{j (\neq i) }[ W_{i \to j}^{(\alpha,+)}+W_{i \to j}^{(\alpha,-)}]$.

In order to measure $P_{S \rightarrow D}(t, t_0)$ one has to detect single directionally-resolved electron transfers between the electrodes and the system and record a sufficiently long time-series of transfer events. Then a histogram of the number of consecutive transfer events $S \rightarrow D$ as function of increasing time intervals ${t-t_0}$ can be generated. By normalizing the histogram by the total number of events in the time series one obtains $P_{S \rightarrow D}(t, t_0)$.

We can calculate the two point-probability in Eq.\.(\ref{lr1}) analytically. We consider the electronic states $\vert 1\rangle$ to $\vert m\rangle$ which belong to charge state $N$ and the electronic states $\vert m+1 \rangle$ to $\vert n\rangle$ that belong to charge state $N+1$. The elementary probability of the $S \rightarrow D$ transfer path is then given by
\begin{equation}\label{plrdecay}
P_{S \rightarrow D}(t,t_0) = \sum_{l=m+1}^n e^{-\gamma_l (t-t_0)} A_{l} B_{l}.
\end{equation}
The decay rates $\gamma_l$ are given by $\gamma_l=\sum_{i=1}^{m} \Gamma^{(D)}_{il} \big(1-f_D(E_{li})\big) + \Gamma^{(S)}_{il} (1-f_S(E_{li})\big)$ only. 
The coefficients $A_{kl}$ are determined by the source and drain couplings and the population of the states:
$A_{l}= \sum_{k=1}^{m} \Gamma^{(D)}_{lk} \big(1-f_D(E_{lk})\big) $ $B_l=\sum_{j=1}^{m} \Gamma^{(S)}_{lj} f_S(E_{lj}) p_j$. The populations are given by the steady state of the system. Surprisingly, each decay rate includes coupling elements to a single state of charge state $N$ only. 

In order to analyze the different time scales in $P_{S \rightarrow D}(t,t_0)$ due to the exponents, we use a one sided inverse Laplace transformation to calculate the decay time distribution $G_{S \rightarrow D}(\gamma)=\int_{0}^{i \infty} \mathrm dt\, e^{(\gamma+i \omega)t}  P_{S \rightarrow D}(t)$. This gives
\begin{equation}
G_{S \rightarrow D}(\gamma)=\sum_{l=m+1}^n  \frac{A_{l} B_{l} }{\gamma_l-\gamma + i \omega}.
\end{equation}
An inverse Laplace transformation over a multiple exponential decays of the form (\ref{plrdecay}) results in a sum over delta-functions. The parameter $\omega$ was introduced for assure a finite width of the peaks in $\vert G_{S \rightarrow D}(\gamma) \vert$. The magnitudes $\vert G_{S \rightarrow D}(\gamma) \vert$ are shown in Fig.\,\ref{fig:5b} for the three electrode configurations. We use a $4.75 V$ bias where a large number of channels is activated. Each peak in Fig.\,\ref{fig:5b} is due to the different time scale occurring in $P_{S \rightarrow D}(t)$. The positions of the peaks are given by the sum over the magnitude of active couplings $\gamma_i$ to a specific electronic state $i$ of the anionic charge state $N+1$ and are identified in Fig.\,\ref{fig:5b}. For all three configurations, only the $\gamma_1$ peak related to the $\Gamma_{11}$ coupling is visible at a bias of $3 V$. Its position is shifted with increasing bias as soon as additional couplings $\Gamma_{j1}$ to state $(1,N)$ are activated. The spectrum of electrode configuration (a) shows dominant peaks for the first three electronic states of the anion. The $\gamma_8$ peak is weak as one could expect by comparing with the spatial profile of $\Gamma_{18}$ in Fig.\ref{fig:1b}. On the contrary a strong $\gamma_8$ peak is found in configuration (b) where the electrodes are in the center of the MgP molecule. Generating a spectrum for different bias and Fermi level by recording the position and the magnitude of the peaks enables one to resolve individual transport channels. The average current solely depends on the sum over the active couplings and one cannot tell them apart. Rastering the molecule with the electrodes would then allow to measure the spatial profiles of the couplings shown in Fig.\ref{fig:1b}. Note that if one applies a high bias to the source electrode, its Fermi function would simplify to $f_S(E_{li})=1$ for all eigenstates and the decay rates $\gamma_l$ would depend on the properties of the coupling to the drain only. In this case the molecule could simply sit on a surface acting as electrode and a tip acting as drain could measure the spatial profile of the couplings. Of interest is also the decay time distribution of transfers in opposite direction of the bias since the sum in $\gamma_l$ over the neutral electronic states would then be replaced by a sum over the anionic states. Combining both spectra would then provide detailed information on each coupling element $\Gamma_{ij}$. Bi-directional electron transfer has been observed in quantum dots\cite{Fuji06}. The practical limitations of the proposed method are set by the inverse Laplace transformation which has to be perfomed numerically for experimental data. 

In summary, the second and the third cumulants provide detailed information and can be used to probe ratios of the couplings between electronic states due to the molecule-electrode bonding. This effect is explained using a simple model. The decay time distribution of two point electron transfer probabilities can resolve active transport channels and the magnitude of the couplings between the electronic states. The prestent methods are general and applicable to different kinds molecules as well as quantum dots.

{\bf Supporting Information Available:} Additional theoretical and 
computational detail are presented.  This material is available free of 
charge via the Internet at http://pubs.acs.org.

{\bf acknowledgments}
We thank Professor Vladimir Chernyak for helpful discussions. The support of the National Science Foundation (Grant No. CHE-0446555, CBC-0533162) and NIRT (Grant No. EEC 0303389)) is gratefully acknowledged. M.E. was partially supported by the FNRS Belgium (charg\'{e} de recherche)


\begin{thebibliography}{10}

\bibitem{Lu03}
W.~Lu,~Z.~J.;\ \ Pfeiï¬er,~L.;\ \ West,~K.~W.;\ \ Rimberg,~A.~J.
  \textit{Nature} \textbf{2003,} \textsl{423,} 422.

\bibitem{Fuji04}
Fujisawa,~T.;\ \ Hayashi,~T.;\ \ Hirayama,~Y.;\ \ Cheong,~H.~D. \textit{Appl.
  Phys. Lett.} \textbf{2004,} \textsl{84,} 2343.

\bibitem{Byla04}
Bylander,~J.;\ \ Duty,~T.;\ \ Delsing,~P. \textit{Nature} \textbf{2005,}
  \textsl{434,} 361.

\bibitem{Gustav06}
Gustavsson,~S.;\ \ Leturcq,~R.;\ \ Simovic,~B.;\ \ Schleser,~R.;\ \ Ihn,~T.;\ \
  Studerus,~P.;\ \ Ensslin,~K.;\ \ Driscoll,~D.~C.;\ \ Gossard,~A.~C.
  \textit{Phys. Rev. Lett.} \textbf{2006,} \textsl{96,} 076605.

\bibitem{Fuji06}
Fujisawa,~T.;\ \ Hayashi,~T.;\ \ Tomita,~R.;\ \ Hirayama,~Y. \textit{Science}
  \textbf{2006,} \textsl{312,} 1634.

\bibitem{Gruneis07}
Gruneis,~A.;\ \ Esplandiu,~M.;\ \ Garcia-Sanchez,~D.;\ \ Bachtold,~A.
  ``"Counting and manipulating single electrons using a carbon nanotube
  transistor"'',  \mbox{cond-mat/0704.1794}.

\bibitem{nitz03a}
Nitzan,~A.;\ \ Ratner,~M.~A. \textit{Science} \textbf{2003,} \textsl{300,}
  1384.

\bibitem{ghos04}
Ghosh,~A.~W.;\ \ Damle,~P.~S.;\ \ Datta,~S.;\ \ Nitzan,~A. \textit{MRS
  Bulletin} \textbf{2004,} \textsl{6,} 391.

\bibitem{levi96}
Levitov,~L.~S.;\ \ ;\ \ Lee,~H.;\ \ Reznikov,~M. \textit{J. Math. Phys.}
  \textbf{1996,} \textsl{37,} 4845.

\bibitem{Bagr03}
Bagrets,~D.;\ \ Nazarov,~Y. \textit{Phys. Rev. B} \textbf{2003,} \textsl{67,}
  085316.

\bibitem{Levi04}
Levitov,~L.~S.;\ \ Reznikov,~M. \textit{Phys. Rev. B} \textbf{2004,}
  \textsl{70,} 115305.

\bibitem{Blant00}
Blanter,~Y.~M.;\ \ Buttiker,~M. \textit{Phys. Rep} \textbf{2000,} \textsl{336,}
  1.

\bibitem{Gurv97}
Gurvitz,~S.~A. \textit{Phys. Rev. B} \textbf{1997,} \textsl{56,} 15215.

\bibitem{Wabnig05}
Wabnig,~J.;\ \ Khomitsky,~D.~V.;\ \ Rammer,~J.;\ \ Shelankov,~A.~L.
  \textit{Phys. Rev. B} \textbf{2005,} \textsl{72,} 165347.

\bibitem{Rammer04}
Rammer,~J.;\ \ Shelankov,~A.~L.;\ \ Wabnig,~J. \textit{Phys. Rev. B}
  \textbf{2004,} \textsl{70,} 115327.

\bibitem{Shela03}
Shelankov,~A.~L.;\ \ Rammer,~J. \textit{Europhys. Lett.} \textbf{2003,}
  \textsl{63,} 485.

\bibitem{Flindt05}
Flindt,~C.;\ \ Novotny,~T.;\ \ Jauho,~A.-P. \textit{Europhys. Lett.}
  \textbf{2005,} \textsl{69,} 475.

\bibitem{Kiess06}
Kiesslich,~G.;\ \ Samuelsson,~P.;\ \ Wacker,~A.;\ \ Schoell,~E. \textit{Phys.
  Rev. B} \textbf{2006,} \textsl{73,} 033312.

\bibitem{Peder05}
Pedersen,~J.~N.;\ \ Wacker,~A. \textit{Phys. Rev. B} \textbf{2005,}
  \textsl{72,} 195330.

\bibitem{groth07}
Groth,~C.;\ \ Michaelis,~B.;\ \ Beenakker,~C. \textit{Phys. Rev. B}
  \textbf{2006,} \textsl{74,} 125315.

\bibitem{max06a}
Espositio,~M.;\ \ Harbola,~U.;\ \ Mukamel,~S. \textit{Phys. Rev. B}
  \textbf{2007,} \textsl{75,} 155316.

\bibitem{sven07_2}
Welack,~S.;\ \ Esposito,~M.;\ \ Harbola,~U.;\ \ Mukamel,~M. ``"Interference effects in the counting statistics of electron transfers through a double quantum dot"'',  \mbox{arXiv:0709.3551v2}.

\bibitem{fleurov07}
Fleurov,~D.;\ \ Eisenberg,~E. ``"Super-Poissonian Shot Noise as a Measure of
  Dephasing in Closed Quantum Dots"'',  \mbox{cond-mat/0705.2668}.

\bibitem{thiel03}
Thielmann,~A.;\ \ Hettler,~M.~H.;\ \ Koenig,~J.;\ \ Schoen,~G. \textit{Phys.
  Rev. B} \textbf{2003,} \textsl{68,} 115105.

\bibitem{thiel05}
Thielmann,~A.;\ \ Hettler,~M.~H.;\ \ Koenig,~J.;\ \ Schoen,~G. \textit{Phys.
  Rev. B} \textbf{2005,} \textsl{71,} 045341.

\bibitem{wang07}
Wang,~S.;\ \ Jiao,~H.;\ \ Li,~F.;\ \ Li,~X.;\ \ Yan,~Y. \textit{Phys. Rev. B}
  \textbf{2007,} \textsl{76,} 125416.

\bibitem{belzig05}
Belzig,~W. \textit{Phys. Rev. B} \textbf{2005,} \textsl{71,} 161301.

\bibitem{Buett05}
Blanter,~Y.~M.;\ \ Buettiker,~M. \textit{Phys. Rev. B} \textbf{1999,}
  \textsl{59,} 10217.

\bibitem{Datta06}
Muralidharan,~B.;\ \ Ghosh,~A.~W.;\ \ Datta,~S. \textit{Phys. Rev. B}
  \textbf{2006,} \textsl{73,} 155410.

\bibitem{Schoell03}
Hettler,~M.;\ \ Wenzel,~W.;\ \ Wegewijs,~M.;\ \ Schoeller,~H. \textit{Phys.
  Rev. Lett.} \textbf{2003,} \textsl{90,} 076805.

\bibitem{Elste05}
Elste,~F.;\ \ Timm,~C. \textit{Phys. Rev. B} \textbf{2005,} \textsl{71,}
  155403.

\bibitem{g03}
Frisch,~M.~J.;\ \ et. al., ``Gaussian 03, \uppercase{R}evision
  \uppercase{C}.02'',  \uppercase{G}aussian, Inc., Wallingford, CT, 2004.

\bibitem{maddox07}
Maddox,~J.~B.;\ \ Harbola,~U.;\ \ Mayoral,~K.;\ \ Mukamel,~S. \textit{J. Phys.
  Chem. C} \textbf{2007,} \textsl{111,} 9516.

\bibitem{gardiner}
Gardiner,~G.;\ \ Zoller,~P. \textit{Quantum Noise;} Springer: 2004.

\bibitem{Muka03}
Mukamel,~S. \textit{Phys. Rev. Lett.} \textbf{2003,} \textsl{90,} 170604.

\end{thebibliography}
\end{document}